\documentclass[sigconf]{acmart}
\copyrightyear{2023}
\acmYear{2023}
\setcopyright{acmlicensed}
\acmConference[SIGIR '23] {Proceedings of the 46th International ACM SIGIR Conference on Research and Development in Information Retrieval}{July 23--27, 2023}{Taipei, Taiwan.}
\acmBooktitle{Proceedings of the 46th International ACM SIGIR Conference on Research and Development in Information Retrieval (SIGIR '23), July 23--27, 2023, Taipei, Taiwan}
\acmPrice{15.00}
\acmISBN{978-1-4503-9408-6/23/07}
\acmDOI{10.1145/XXXXXX.XXXXXX}

\usepackage{multirow}

\begin{document}

\title{LADER: Log-Augmented DEnse Retrieval for Biomedical Literature Search}


\author{Qiao Jin}
\affiliation{%
  \institution{National Institutes of Health}
  \city{Bethesda}
  \state{Maryland}
  \country{USA}}
\email{qiao.jin@nih.gov}

\author{Ashley Shin}
\affiliation{%
  \institution{National Institutes of Health}
  \city{Bethesda}
  \state{Maryland}
  \country{USA}}
\email{ashley.shin@nih.gov}

\author{Zhiyong Lu}
\affiliation{%
  \institution{National Institutes of Health}
  \city{Bethesda}
  \state{Maryland}
  \country{USA}}
\email{zhiyong.lu@nih.gov}

\renewcommand{\shortauthors}{Jin, et al.}

\begin{abstract}
Queries with similar information needs tend to have similar document clicks, especially in biomedical literature search engines where queries are generally short and top documents account for most of the total clicks.
Motivated by this, we present a novel architecture for biomedical literature search, namely Log-Augmented DEnse Retrieval (LADER), which is a simple plug-in module that augments a dense retriever with the click logs retrieved from similar training queries.
Specifically, LADER finds both similar documents and queries to the given query by a dense retriever.
Then, LADER scores relevant (clicked) documents of similar queries weighted by their similarity to the input query.
The final document scores by LADER are the average of (1) the document similarity scores from the dense retriever and (2) the aggregated document scores from the click logs of similar queries.
Despite its simplicity, LADER achieves new state-of-the-art (SOTA) performance on TripClick, a recently released benchmark for biomedical literature retrieval.
On the frequent (``HEAD'') queries, LADER largely outperforms the best retrieval model by 39\% relative NDCG@10 (0.338 v.s. 0.243).
LADER also achieves better performance on the less frequent (``TORSO'') queries with 11\% relative NDCG@10 improvement over the previous SOTA (0.303 v.s. 0.272).
On the rare (``TAIL'') queries where similar queries are scarce, LADER still compares favorably to the previous SOTA method (NDCG@10: 0.310 v.s. 0.295).
On all queries, LADER can improve the performance of a dense retriever by 24\%-37\% relative NDCG@10 while not requiring additional training,
and further performance improvement is expected from more logs.
Our regression analysis has shown that queries that are more frequent, have higher entropy of query similarity and lower entropy of document similarity, tend to benefit more from log augmentation.
\end{abstract}





\maketitle

\section{Introduction}
\begin{figure}[h]
  \centering
  \includegraphics[width=0.94\linewidth]{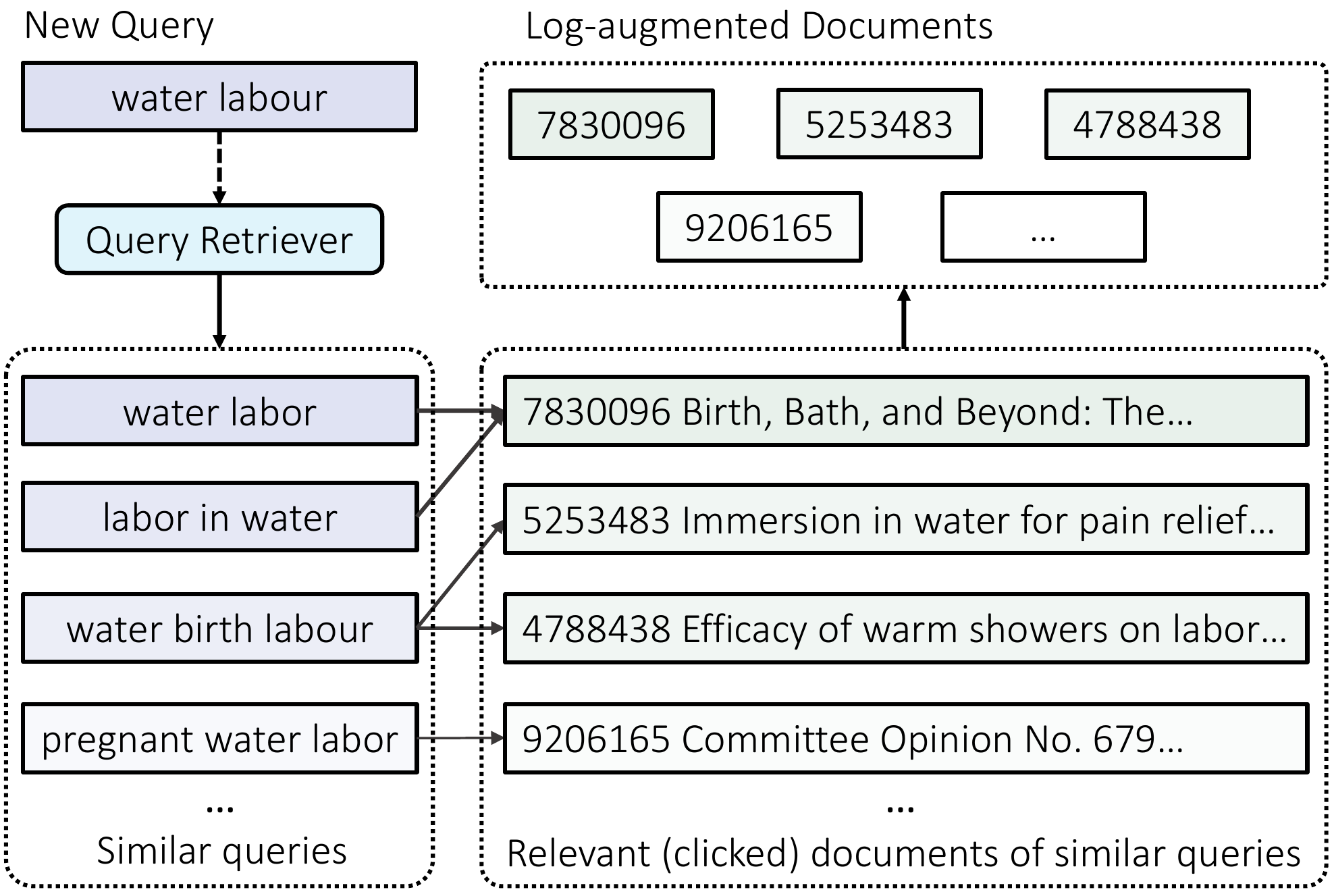}
  \caption{An example of using click logs of retrieved similar queries to augment biomedical literature search.}
  \Description{An example of using click logs of retrieved similar queries to augment biomedical literature search.}
  \label{fig:example}
\end{figure}

Biomedical literature search is an essential step for knowledge discovery and clinical decision support \cite{DBLP:journals/jbi/GopalakrishnanJ19, DBLP:journals/jamia/ElyOCER05, jin2022state}.
It has several distinct properties from other information retrieval (IR) tasks:
(1) Most queries are short. The average length of a query is about 3.5 tokens in PubMed\footnote{\url{https://pubmed.ncbi.nlm.nih.gov/}}, a widely used biomedical literature search engine \cite{dogan2009log, fiorini2018best, fiorini2018user};
(2) Large-scale relevant query-document pairs can be easily collected from the user click logs.
TripClick \cite{rekabsaz2021tripclick}, a recently released benchmark for biomedical literature retrieval, contains 1.3 million query-document relevance signals collected from the Trip Database\footnote{\url{https://www.tripdatabase.com/}} logs;
(3) Users mostly browse the documents on the first page \cite{dogan2009log}, and the top 31\% most clicked documents account for 80\% clicks in the released Trip Database logs.
These unique characteristics motivate the augmentation of biomedical literature search by directly retrieving from the click logs of similar queries, since queries with similar information needs tend to have similar document clicks \cite{DBLP:journals/jasis/Baeza-YatesHM07, DBLP:conf/ecir/YinSC09, DBLP:conf/sigir/ZhuangLZ22}.
This approach is essentially similar to recent retrieval augmentation methods \cite{DBLP:conf/iclr/KhandelwalLJZL20, frisoni-etal-2022-bioreader, DBLP:journals/corr/abs-2303-00534}.
Fig \ref{fig:example} shows an example query and clicked documents from its similar queries in the training set of TripClick.
The objective of this work is to augment biomedical literature search with such documents.

One model can simply return clicked documents from the logs of similar queries as the retrieval results, which is analogous to memory-based collaborative filtering in recommendation systems \cite{DBLP:journals/advai/SuK09}.
However, this might miss relevant documents that have not been clicked before.
Therefore, clicked documents from logs are better used to augment an existing retriever, and they are theoretically retriever-agnostic.
We choose to use dense retrievers, where queries and documents are encoded and matched in low-dimensional embedding space, because recent work has shown that dense retrievers based on pre-trained language models such as BERT \cite{DBLP:conf/naacl/DevlinCLT19} outperform traditional sparse retrievers on various tasks \cite{DBLP:conf/emnlp/KarpukhinOMLWEC20, DBLP:conf/sigir/KhattabZ20, DBLP:conf/iclr/XiongXLTLBAO21}.

In this paper, we introduce Log-Agumented DEnse Retrieval (LADER), which is a novel and simple plug-in module that augments dense retrievers by interpolating the document scores aggregated from logs of similar queries and the original scores from a dense retriever.
We conduct experiments on the recently introduced TripClick benchmark \cite{rekabsaz2021tripclick} and show that: (1) LADER outperforms previous state-of-the-art (SOTA) models on queries of all frequency groups, with close to 40\% relative NDCG@10 improvement on the frequent (``HEAD'') queries; (2) LADER improves the backbone dense retriever by 24\%-37\% relative NDCG@10 while not requiring additional training, and it is expected to have further performance improvements with more logs.
Our regression analysis shows that queries with higher frequency, lower entropy of similar query scores, and higher entropy of similar document scores, tend to benefit more from log augmentation.

\section{Methods}
We describe the training of our backbone dense retriever in \S\ref{dr}, and introduce how to do inference with the LADER module in \S\ref{LADER}.

\subsection{Dense Retriever Training} \label{dr}
We train a dense retriever that contains a query encoder ($\text{QEnc}$) and a document encoder ($\text{DEnc}$), both are 12-layer transformer ($\text{Trm}$) encoders \cite{DBLP:conf/nips/VaswaniSPUJGKP17}.
The encoders are initialized with PubMedBERT-base \cite{gu2021pubmedbert}, which is a biomedical domain-specific BERT model.
During training, each instance contains a triple of a query $q$, a relevant document $d^{+}$ for the query, and a non-relevant document $d^{-}$ for the query.
They are fed to their respective encoders and the last $\text{[CLS]}$ hidden states are used as their embeddings:
\[E(q) = \text{QEnc}(q) = \text{Trm}(\text{[CLS]}\ q\ \text{[SEP]})_\text{[CLS]}\]
\[E(d) = \text{DEnc}(d) = \text{Trm}(\text{[CLS]}\ d_\text{title}\ \text{[SEP]}\ d_\text{abstract}\ \text{[SEP]})_\text{[CLS]}\]
where $\text{[CLS]}$ and $\text{[SEP]}$ are special tokens used in BERT.

We optimize the model with a combination of two loss functions: an in-batch negative log-likelihood loss $\mathcal{L}^{i}$ \cite{DBLP:conf/emnlp/KarpukhinOMLWEC20} and a triplet contrastive loss  $\mathcal{L}^{t}$ \cite{DBLP:conf/cvpr/SchroffKP15}.
The in-batch negative log-likelihood loss helps the dense retriever distinguish between positive documents and all other documents in the mini-batch:
\[\mathcal{L}^{i} = -\text{log} \frac{\text{exp}(E(q)^T E(d^{+}))}{\sum_{i \in B} \text{exp}(E(q)^T E(d_i))}\]
where $B$ denotes all documents from the mini-batch.
The triplet loss further contrasts positive documents and hard negative documents:
\[\mathcal{L}^{t} = \text{max}(0, \text{Dist}(E(q), E(d^{+})) - \text{Dist}(E(q), E(d^{-}) + \alpha)\]
where $\text{Dist}$ denotes a distance metric and $\alpha$ is the target margin between the positive and negative pairs.

The final loss is a weighted sum of the two loss functions:
\[\mathcal{L} = \beta\mathcal{L}^{i} + (1 - \beta)\mathcal{L}^{t}\]
where $\beta$ is a hyper-parameter for loss weighting. We train parameters in $\text{QEnc}$ and $\text{DEnc}$ end-to-end by gradient-based optimizers.

\begin{table*}[hbt!]
  \caption{Model performance on the TripClick test sets. Baseline results are from \cite{rekabsaz2021tripclick}. LA: log augmentation; DR: dense retrieval.}
  \label{tab:main}
  \begin{tabular}{lccccccccc}
    \toprule
    \multirow{2}{*}{\textbf{Model}} & \multicolumn{3}{c}{\textbf{HEAD} (\textbf{DCTR} relevance)} & \multicolumn{3}{c}{\textbf{TORSO} (\textbf{RAW} relevance)} & \multicolumn{3}{c}{\textbf{TAIL} (\textbf{RAW} relevance)} \\
    \cmidrule(r){2-4} \cmidrule(r){5-7} \cmidrule(r){8-10} & \textbf{NDCG@10} 
    & \textbf{MRR} & \textbf{Recall@10} & \textbf{NDCG@10} & \textbf{MRR} & \textbf{Recall@10} & \textbf{NDCG@10} & \textbf{MRR} & \textbf{Recall@10} \\ 
    \midrule
    BM25 \cite{DBLP:journals/ftir/RobertsonZ09} & 0.140 & 0.290 & 0.138 & 0.206 & 0.283 & 0.262 & 0.267 & 0.258 & 0.409 \\
    RM3 PRF \cite{DBLP:conf/sigir/LavrenkoC01, DBLP:conf/cikm/LvZ09a} & 0.141 & 0.300 & 0.136 & 0.194 & 0.261 & 0.254 & 0.242 & 0.227 & 0.384 \\
    PACRR \cite{DBLP:conf/emnlp/HuiYBM17} & 0.175 & 0.356 & 0.162 & 0.212 & 0.302 & 0.262 & 0.267 & 0.261 & 0.409 \\
    MP \cite{DBLP:conf/aaai/PangLGXWC16} & 0.183 & 0.372 & 0.173 & 0.243 & 0.347 & 0.297 & 0.281 & 0.280 & 0.409 \\
    KNRM \cite{DBLP:conf/sigir/XiongDCLP17} & 0.191 & 0.393 & 0.173 & 0.235 & 0.338 & 0.283 & 0.272 & 0.265 & 0.409 \\
    ConvKNRM \cite{DBLP:conf/wsdm/DaiXC018} & 0.198 & 0.420 & 0.178 & 0.243 & 0.358 & 0.288 & 0.271 & 0.265 & 0.409 \\
    TK \cite{DBLP:conf/sigir/HofstatterZMCH20} & 0.208 & 0.434 & 0.189 & 0.272 & 0.381 & 0.321 & 0.295 & 0.279 & \textbf{0.459} \\
    \midrule
    LADER (ours) & \textbf{0.338} & \textbf{0.664} & \textbf{0.304} & \textbf{0.303} & \textbf{0.427} & \textbf{0.353} & \textbf{0.310} & \textbf{0.306} & 0.449 \\
    LADER w/o LA & 0.247 & 0.532 & 0.237 & 0.241 & 0.350 & 0.293 & 0.260 & 0.257 & 0.394 \\
    LADER w/o DR & 0.324 & 0.649 & 0.284 & 0.266 & 0.396 & 0.298 & 0.232 & 0.236 & 0.330\\
    \bottomrule
  \end{tabular}
\end{table*}

\subsection{LADER Inference} \label{LADER}
\begin{figure}[h]
  \centering
  \includegraphics[width=\linewidth]{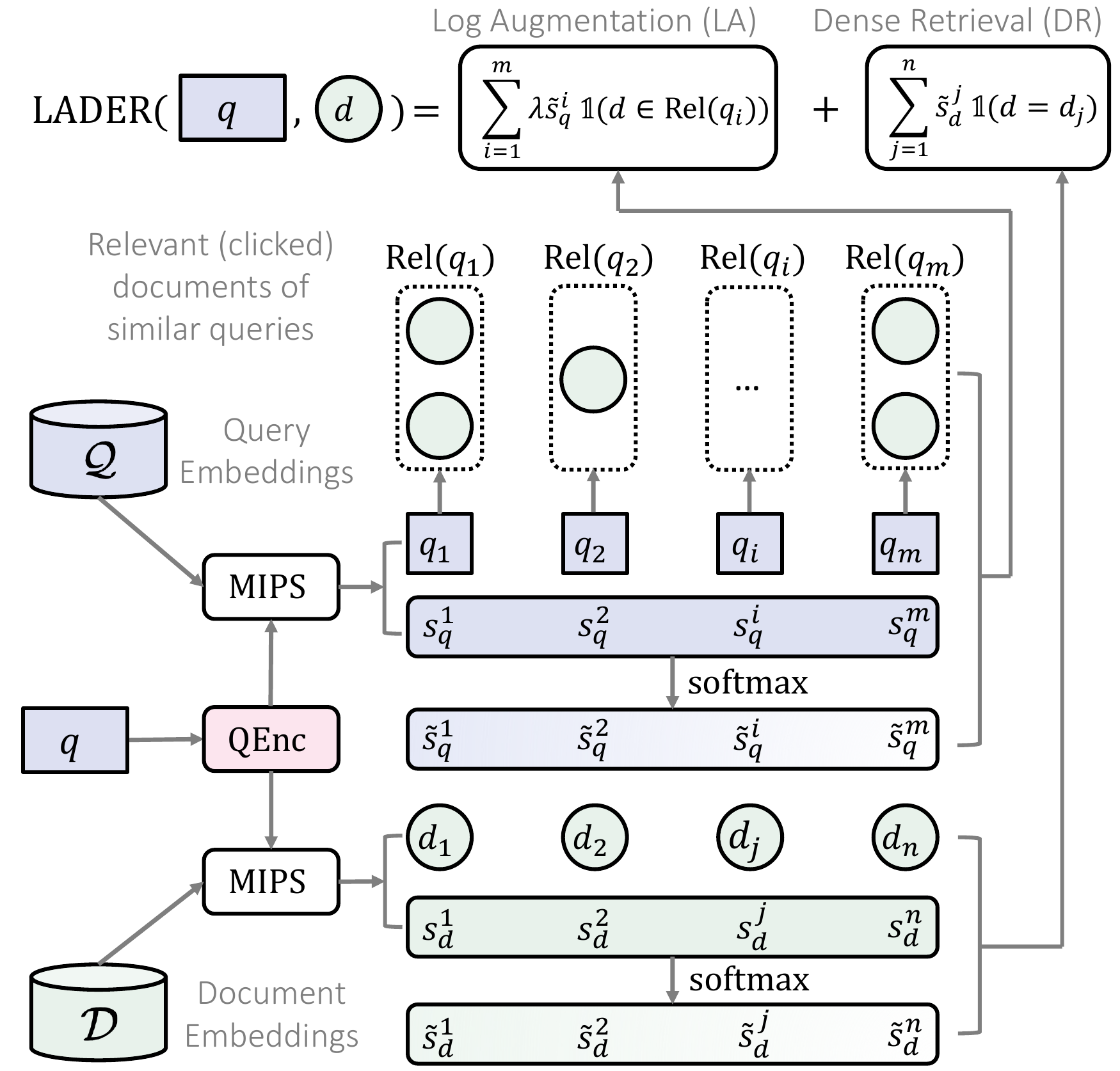}
  \caption{Overall architecture of the LADER model.}
  \Description{Overall architecture of the LADER model.}
  \label{fig:arch}
\end{figure}
The overall architecture of LADER is shown in Figure~\ref{fig:arch}.
We use the trained $\text{QEnc}$ to encode all queries in the training dataset and use the trained $\text{DEnc}$ to encode all documents in the collection, getting $\mathcal{Q} \in \mathbb{R}^{N_q \times 768}$ and $\mathcal{D} \in \mathbb{R} ^ {N_d \times 768}$, respectively. $N_q$ denotes the number of training queries and $N_d$ denotes the number of documents in the collection.

During inference, we first encode a new query $q$ by $\text{QEnc}$. Then, we conduct a maximum inner-product search ($\text{MIPS}$) to get the top-$m$ most similar queries ($q_1, q_2, ..., q_m$) and their inner-product similarities ($s_q^1, s_q^2, ..., s_q^m$) from the training set, as well as the top-$n$ most similar documents ($d_1, d_2, ..., d_n$) and their similarities ($s_d^1, s_d^2, ..., s_d^n$) from the collection:
\[(q_1, s_q^1), (q_2, s_q^2), ..., (q_m, s_q^m) = \text{MIPS}(\text{QEnc}(q), \mathcal{Q})\]
\[(d_1, s_d^1), (d_2, s_d^2), ..., (d_n, s_d^n) = \text{MIPS}(\text{QEnc}(q), \mathcal{D})\]
The similarity scores are further normalized by a softmax function:
\[\tilde{s}_q^1, \tilde{s}_q^2, ..., \tilde{s}_q^m = \text{softmax}(s_q^1, s_q^2, ..., s_q^m)\]
\[\tilde{s}_d^1, \tilde{s}_d^2, ..., \tilde{s}_d^n = \text{softmax}(s_d^1, s_d^2, ..., s_d^n)\]
We denote the mapping from a query to its relevant documents in the training set as $\text{Rel}: q \mapsto \{d\}$.
The final score of a document $d$ in the candidate set $\{d_1, d_2, ..., d_n\} \cup \{\text{Rel}(q_1), \text{Rel}(q_2), ..., \text{Rel}(q_m)\}$ is a weighted sum computed as follows:
\[\text{Score}(d) = \sum_{j \in [1, ..., n]} \tilde{s}_d^j\ \mathbf{1}(d = d_j) + \sum_{i \in [1, ..., m]} \lambda \tilde{s}_q^i\ \mathbf{1}(d \in \text{Rel}(q_i)) \]
where the former part is the dense retrieval score and the latter part is the log-augmentation score from similar queries, $\mathbf{1}$ is the indicator function, and $\lambda$ is a hyper-parameter to control the extent of log augmentation.
We return the candidate documents ranked by their final scores to the input query.


\section{Experiments}

\subsection{Settings}
\paragraph{\textbf{Dataset.}}
We evaluate our LADER method on the TripClick benchmark \cite{rekabsaz2021tripclick}, which contains 692k unique queries and 1.5M documents (PubMed abstracts).
Based on their frequencies, the queries are divided into HEAD ($>$44), TORSO (6--44), and TAIL ($<$6) subsets.
Both the validation and test sets contain 1,175 queries for each HEAD, TORSO and TAIL subset.
Following \cite{rekabsaz2021tripclick, DBLP:conf/ecir/HofstatterASH22}, we use two sets of relevance scores:
The ``RAW'' relevance is used to judge TORSO and TAIL queries, where clicked documents have a score of 1 and other documents have a score of 0.
The Graded Document Click-Through Rate \cite{DBLP:series/synthesis/2015Chuklin, DBLP:conf/wsdm/CraswellZTR08} (``DCTR'') relevance is used to judge the HEAD queries, where the document score is defined as the number of clicks divided by the number of exposure.
To train the dense retriever (\S\ref{dr}), we use 10M triples released by \cite{DBLP:conf/ecir/HofstatterASH22}, where each triple contains a TripClick training query, a clicked document, and a non-relevant document sampled from BM25 negatives.

\paragraph{\textbf{Configuration.}}
We implement LADER with PyTorch \cite{DBLP:conf/nips/PaszkeGMLBCKLGA19} and HuggingFace's libraries \cite{wolf-etal-2020-transformers}.
For training the dense retriever, we use the AdamW optimizer \cite{DBLP:journals/corr/KingmaB14, DBLP:conf/iclr/LoshchilovH19} for 20k steps with the learning rate of 2e-5 and weight decay of 1e-2, batch size of 256, 10k warmup steps, the cosine learning rate decay schedule, $\alpha=5.0$, $\beta=0.9$, and the Euclidean distance for the triplet loss.
For inference, we implement MIPS with FAISS's \texttt{FlatIP} index \cite{johnson2019billion}, $m=n=1000$, $N_q=685,649$, and $N_d=1,523,871$.
We use $\lambda=0.5$ for the HEAD and TORSO queries, and $\lambda=0.2$ for the TAIL queries.
We also experiment with two ablations: LADER w/o Log Augmentation (LA) where LA scores are set to 0, and LADER w/o Dense Retrieval (DR) where DR scores are set to 0.

\paragraph{\textbf{Baseline methods for comparison.}}
We compare LADER with various baselines in \cite{rekabsaz2021tripclick} on all queries, including BM25 \cite{DBLP:journals/ftir/RobertsonZ09}, RM3 Pseudo Relevance Feedback (PRF) \cite{DBLP:conf/sigir/LavrenkoC01, DBLP:conf/cikm/LvZ09a}, Position Aware Convolutional Recurrent Relevance Matching (PACRR) \cite{DBLP:conf/emnlp/HuiYBM17}, Match Pyramid (MP) \cite{DBLP:conf/aaai/PangLGXWC16}, Kernel-based Neural Ranking Model (KNRM) \cite{DBLP:conf/sigir/XiongDCLP17}, Convolutional KNRM (ConvKNRM) \cite{DBLP:conf/wsdm/DaiXC018}, and Transformer-Kernel (TK) \cite{DBLP:conf/sigir/HofstatterZMCH20}.
For the HEAD queries, we further compare with current SOTA methods \cite{DBLP:conf/ecir/HofstatterASH22} based on pre-trained language models, including bi-encoders initialized by different BERT models \cite{DBLP:journals/corr/abs-1910-01108, DBLP:conf/emnlp/BeltagyLC19, gu2021pubmedbert}.
Following their respective metrics, we compare with benchmark baselines \cite{rekabsaz2021tripclick} using NDCG@10, MRR, and Recall@10, and compare with HEAD query SOTA \cite{DBLP:conf/ecir/HofstatterASH22} using NDCG@10, MRR@10, and also Recall@1k.

\subsection{Main Results}

\paragraph{\textbf{Comparison with benchmark baselines}}
Table~\ref{tab:main} shows comprehensive comparisons of LADER with a variety of benchmark baselines \cite{rekabsaz2021tripclick} on all queries.
LADER outperforms all benchmark baselines on each query subset, and on all metrics except the Recall@10 for the TAIL queries.
On the most frequent HEAD queries, LADER outperforms the best benchmark baseline method (TK) by large margins, showing 63\% (0.338 v.s. 0.208), 53\% (0.664 v.s. 0.434), and 61\% (0.304 v.s. 0.189) relative gains on NDCG@10, MRR, and Recall@10, respectively.
On the less frequent TORSO queries, the performance improvements over the previous SOTA are decent with 10\% to 12\% relative gains on all metrics.
On the rare TAIL queries, LADER still performs favorably than previous SOTA on NDCG@10 (0.310 v.s. 0.295) and MRR (0.306 v.s. 0.280), but the performance is slightly lower on Recall@10 (0.449 v.s. 0.459).

\paragraph{\textbf{Comparison with BERT DOT}}
In Table~\ref{tab:head}, we compare LADER with more recent BERT-based retrievers (BERT DOT) \cite{DBLP:conf/ecir/HofstatterASH22}, whose results are only available on the HEAD queries. 
Compared with the BERT DOT dense retriever, LADER has 39\% (0.338 v.s. 0.243), 24\% (0.659 v.s. 0.530), and 7\% (0.893 v.s. 0.828) relative improvement on NDCG@10, MRR@10, and Recall@1k, respectively.
This shows the effectiveness of log augmentation since our backbone dense retriever (LADER w/o LA) performs similarly to their counterparts.

\paragraph{\textbf{Improvement over the backbone dense retriever}}
On all queries, LADER improves the performance of the backbone dense retriever by 24\%-37\% relative NDCG@10 (LADER v.s. LADER w/o LA) while not requiring additional training.
The performance improvement is more significant on the HEAD queries than on the TORSO or TAIL queries, which will be further analyzed in the next section.

\begin{table}
  \caption{LADER results on the HEAD queries compared to BERT DOT \cite{DBLP:conf/ecir/HofstatterASH22}; LA: log augmentation; DR: dense retrieval.}
  \label{tab:head}
  \begin{tabular}{lccc}
    \toprule
    \textbf{Model} & \textbf{NDCG@10} & \textbf{MRR@10} & \textbf{R@1k} \\
    \toprule
    BM25 \cite{DBLP:journals/ftir/RobertsonZ09} & 0.140 & 0.276 & 0.834  \\
    \multicolumn{4}{l}{BERT DOT
    \cite{DBLP:conf/ecir/HofstatterASH22}}  \\
    \hspace{0.1cm} w/ DistillBERT \cite{DBLP:journals/corr/abs-1910-01108} & 0.236 & 0.512 & 0.813 \\
    \hspace{0.1cm} w/ SciBERT \cite{DBLP:conf/emnlp/BeltagyLC19} & 0.243 & 0.530 & 0.793 \\
    \hspace{0.1cm} w/ PubMedBERT \cite{gu2021pubmedbert} & 0.235 & 0.509 & 0.828 \\
    \midrule
    LADER (ours) & \textbf{0.338} & \textbf{0.659} & \textbf{0.893} \\
    LADER w/o LA & 0.247 & 0.526 & 0.878 \\
    LADER w/o DR & 0.324 & 0.644 & 0.670 \\
    \midrule
    \multicolumn{4}{l}{Log-Augmented (LA-)}  \\
    \hspace{0.1cm} Sparse retriever (BM25) & 0.312 & 0.598 & 0.889 \\
    \hspace{0.1cm} Raw PubMedBERT \cite{gu2021pubmedbert} & 0.137 & 0.288 & 0.453\\
    \bottomrule
  \end{tabular}
\end{table}

\subsection{Analysis}

\paragraph{\textbf{Importance of trained dense retrievers.}}
In Table~\ref{tab:head}, we show the results of log augmentation with different backbone retrievers.
Log-augmented raw PubMedBERT performs much worse than LADER (0.137 v.s. 0.338 NDCG@10), suggesting the importance of the backbone dense retriever to be trained on retrieval tasks (\S\ref{dr}).
We also replace the dense retriever in LADER with BM25, which we denote as LABM25 and implement with Pyserini \cite{Lin_etal_SIGIR2021_Pyserini}.
LABM25 greatly improves the original BM25 baseline by 122\% (0.312 v.s. 0.140 NDCG@10), which proves that the effectiveness of log augmentation is retriever-agnostic.
However, LADER still outperforms LABM25 (0.338 v.s. 0.312 NDCG@10), indicating that the potential of log augmentation is better harnessed by dense retrievers.

\begin{figure}[h]
  \centering
  \includegraphics[width=\linewidth]{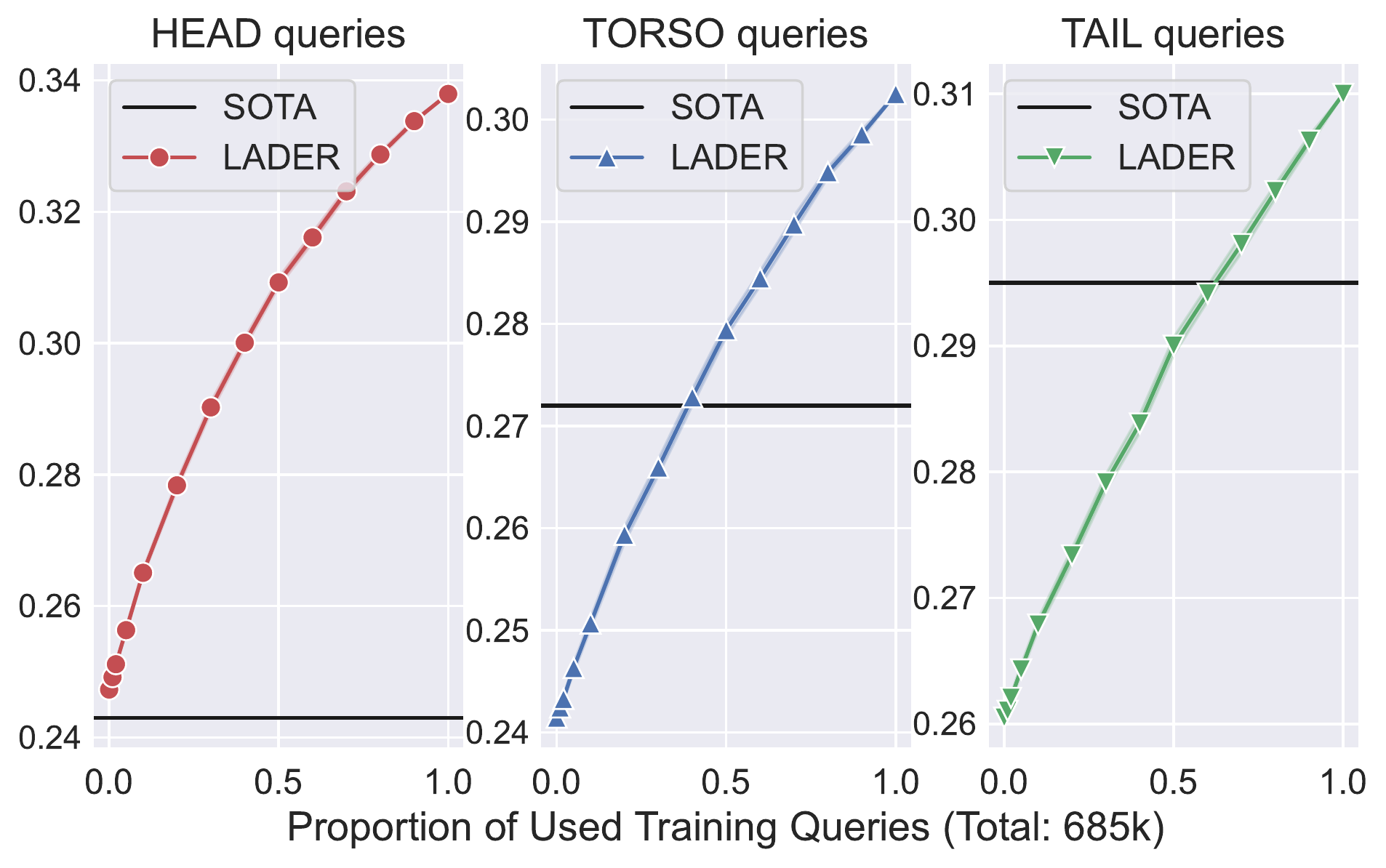}
  \caption{NDCG@10 of LADER with different proportions of training queries to retrieve in log-augmentation.}
  \Description{NDCG@10 of LADER with different proportions of training queries to retrieve in log-augmentation.}
  \label{fig:proportion}
\end{figure}

\paragraph{\textbf{Effects of log size.}}
In Figure~\ref{fig:proportion}, we show the performance of LADER using different proportions of queries sampled from the training set for the log augmentation.
For all queries, the performance improves with the number of queries to retrieve, and since these curves have not saturated yet, more performance gains are expected with more logs.
We also find that on lower frequency queries, more training queries are required to retrieve from for LADER to outperform current SOTA methods.

\begin{figure}[h]
  \centering
  \includegraphics[width=\linewidth]{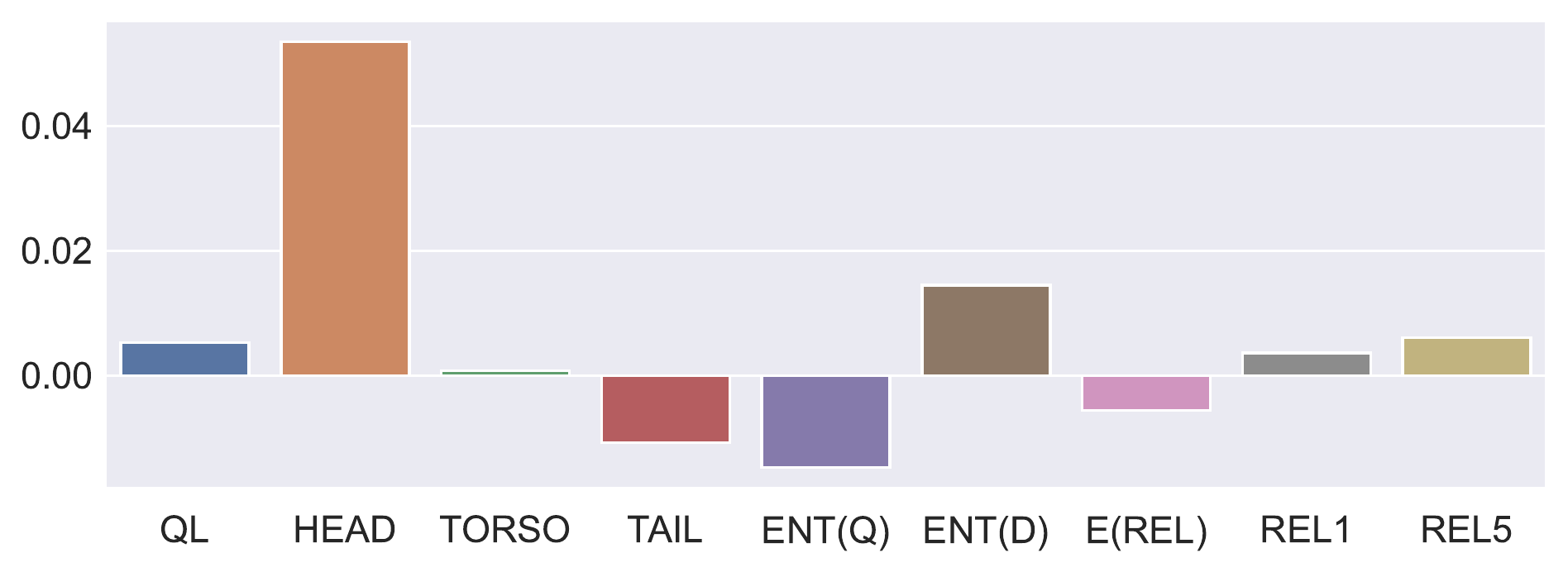}
  \caption{Feature coefficients in the regression analysis. Positive values indicate more gains from log augmentation.}
  \Description{Feature coefficients in the regression analysis.}
  \label{fig:feature}
\end{figure}

\paragraph{\textbf{What queries gain more from log-augmentation?}}
We collect 9 features for each query, including query length (QL), query group (HEAD, TORSO or TAIL), ENT(Q): the entropy of $[\tilde{s}_q^1, \tilde{s}_q^2, ..., \tilde{s}_q^m]$, ENT(D): the entropy of $[\tilde{s}_d^1, \tilde{s}_d^2, ..., \tilde{s}_d^n]$, E(REL): the expectation of relevant document number, and the average number of relevant documents in the top-1 and top-5 similar queries (REL1 and REL5).
We fit a linear regression model to predict the gain of NDCG@10 by log-augmentation (LADER v.s. LADER w/o LA) using min-max normalized query features.
The feature coefficients shown in Figure~\ref{fig:feature} indicate that:
(1) Query frequency is the most important feature: more frequent (HEAD group) queries benefit more than average from log augmentation, while rare (TAIL group) queries benefit less than average;
(2) Queries without very similar documents (higher ENT(D)), benefit more than queries with very similar documents;
(3) Queries with very similar queries in the log (lower ENT(Q)), benefit more from log augmentation than queries without.

\section{Conclusions and Limitations}
We present LADER, a simple and novel plug-in module that uses search logs to augment dense retrievers.
Our results show that LADER achieves new SOTA on TripClick, and can largely improve the performance of its backbone retriever without additional training.
We also provide thorough analyses of its characteristics.

One limitation of LADER is that it increases the search latency by about $N_q / N_d$ (45\% for TripClick) due to the additional query-to-query retrieval step.
Another drawback of this study is that we only use data from one search engine.
It remains interesting to test the generalizability of LADER to a different search engine, which will be beneficial for cold-starting new literature search initiatives.

\begin{acks}
We are grateful to the TripClick benchmark organizers for sharing the data. We also thank the SIGIR reviewers for their constructive comments. This research was supported by the NIH Intramural Research Program, National Library of Medicine.
\end{acks}

\bibliographystyle{ACM-Reference-Format}
\bibliography{reference}


\end{document}